\documentstyle[preprint,aps,prc,epsf]{revtex}

\newcommand{\be}{\begin{equation}}
\newcommand{\ee}{\end{equation}}
\newcommand{\bea}{\begin{eqnarray}}
\newcommand{\eea}{\end{eqnarray}}

\begin{document}
\draft
\title{Pions in the nuclear medium and Drell-Yan scattering}
\author{A.~E.~L.~Dieperink}
\address{Kernfysisch Versneller Instituut, Zernikelaan 25, NL-9747AA
Groningen, The Netherlands}
\author{C.~L.~Korpa}
\address{Department of Theoretical Physics, Janus Pannonius
University, Ifjusag u.\ 6, 7624 Pecs, Hungary}
\date{\today}
\maketitle

\begin{abstract}
We investigate the modification of the
pion-cloud
in the
nuclear medium and its effect on the nuclear Drell-Yan process.
The pion's in-medium self-energy is calculated in a
self-consistent delta-hole model, with particle-hole contribution
also included. Both the imaginary and real part of the pion's
and delta's
self-energy are taken into account and related through a
dispersion relation assuring causality. The resulting in-medium
pion light-cone momentum
distribution shows only a slight enhancement compared to
the one of the free nucleon.
As a consequence the ratio of the cross-section for Drell-Yan
scattering on nuclear
matter and nucleonic target
is close to unity in agreement with experiment.
\end{abstract}


\section{Introduction}
Recently there has been renewed interest in the role of the pion
propagator in the nuclear medium for several reasons:
First the ratio of spin-longitudinal  and spin-transverse response
functions   below
the quasi-elastic peak in a naive  model is predicted
to be much larger than unity, while experiment finds
a ratio close to unity in $(\vec{p},\vec{n})$ polarization transfer
\cite{Taddeucci}.
Second the  observed ratio of cross sections for Drell-Yan
scattering on nuclear and nucleonic targets is consistent
with no excess pions present in the nucleus,
i.e. no enhancement of the sea-quarks \cite{Alde}.

In the past the effect of medium modifications of the
pion propagator has been investigated in detail by Ericson
and Thomas \cite{Thomas} (in connection with the EMC effect)
and Bickerstaff et al.\ \cite{Bicker}.
The apparent absence of the pion-cloud enhancement in DY scattering
was explained by Brown et al.\ \cite{Brown} in terms of
partial restoration of chiral symmetry and the associated decrease
in masses of the nucleons and vector mesons in the nuclear medium.
It was also pointed out \cite{Szczurek} that, in a definite model
considered, the correct normalization of the  physical
nucleon state, consisting of the bare nucleon and pion-cloud term,
considerably reduces the effect of the pion-cloud enhancement.

The sensitivity of the pion-cloud to the pion-nucleon-nucleon
($\pi NN$) vertex cut-off $\Lambda$ was emphasized by Thomas
\cite{Thomas1}. Not only is
the sea-quark content of the nucleon very sensitive
to the value of the cut-off, the
enhancement of the pion light-cone-momentum distribution
in the nuclear medium also
follows that pattern, even if the value of the cut-off
is kept the same in the medium as in free space \cite{Bick}.
Another uncertainty concerns the values of the Migdal $g'$
parameters ($g'_{NN}, g'_{N\Delta},g'_{\Delta\Delta}$),
describing the effects of short-range repulsion. Taking
large enough values of these parameters it is possible to
reduce the medium enhancement of the pion distribution.
However, it has not been reliably established that using
such large values of the $g'$ parameters is physically
justified.

Our approach in reexamining the pion-cloud enhancement
is to perform a more ambitious
computation of the pion
light-cone-momentum distribution in the nuclear medium,
using the energy and momentum dependent pion self-energy
and propagator. The latter are calculated in a recently
developed self-consistent delta-hole model, including particle-hole
states, and allowing for a width of the delta
\cite{Korpa}. This model takes
into account both the real and the imaginary parts of the
pion's and delta's self-energy, in a way that assures causality
and absence of
unphysical poles. The Schwinger-Dyson equations (without vertex
corrections) for the delta and the pion are solved
self-consistently, while the nucleon is treated in a mean-field
approximation.

The framework of the present approach (and that of Ref.\ 
\cite{Korpa}) is the effective
field theory of hadrons, which is used consistently to calculate
physical processes. The basic difference with previous approaches
is the self-consistent treatment of the delta-hole contribution
to the pion self-energy, including both its real and imaginary
part. As already noted in Ref.\ \cite{Thomas} much of the
pion-distribution enhancement came from the delta-hole term
through its coupling to particle-hole term by $g'_{N\Delta}$.
However, that result was based on a crude approximation for the
delta-hole contribution, including only its real part
(violating causality), with a momentum dependence coming only
from the p-wave nature of the coupling and the form factor.

For the practical calculation of the pion light-cone
momentum distribution in isospin symmetric nuclear matter
(at temperature $T=0$) we compare two approaches.
First in section II we present a calculation of the pion
light-cone-momentum distribution, based on explicit
summation of the relevant diagrams involving the dressed
pion propagator, and using an explicit integration over the
initial and final nucleon momenta.

Second, in section III we consider the computation
of the same quantity, but based on the nuclear-matter response
function. While somewhat simpler, this approach gives only
an approximate (however, for typical nuclear densities very close)
expression for the pion distribution. This happens since the
calculation of the imaginary part of the pion self-energy
involves the full phase-space factors for the two
nucleons \cite{Veltman}, while such momentum dependent
factor is not present for the incoming nucleon in the calculation
involving explicit summation of relevant diagrams.
However, the response-function approach takes
into account also the process when a nucleon becomes
a delta plus a pion, if the pion self-energy includes the
delta-hole contribution. We checked numerically that the
two approaches give practically the same result.
The ratio for Drell-Yan scattering
on an isospin-symmetric nuclear target and the deuteron
is presented in section IV. Numerical results, a
discussion and comparison with other computations and
experimental results is the subject of section V.

\section{Pion distribution in the nuclear medium}
It was noticed long time ago that the pion cloud gives a
scaling contribution to the deep inelastic lepton scattering
on the free nucleon \cite{Sullivan}.
The nucleon wave function can be schematically expressed as
\be |N>_{\rm {phys}} =\sqrt{Z} |N>_{\rm{bare}} +\alpha|N\pi>
   + \beta |\Delta\pi>+\cdots \ee
In this approach the  light-cone momentum distribution of a
quark with flavor $f$
in a proton can be written as ($B=N,\Delta$)
\be q_f(x)= Z q_{f,bare}(x) +
\sum_{B,i} c_i\left[ \int_x^1 \frac{dy}{y} f^{B_i/N}(y)
q^{B_i}_{f,bare}(x/y)
 + \int^1_x \frac{dy}{y} f^{\pi_i/N}(y) q^{\pi_i}(x/y)\right] .
\label{pphys}\ee
Here $c_i$ ($i$ labels the charge states)
are the appropriate isospin Clebsch-Gordan coefficients.
Note that  following \cite{MT} we  use the $Z$ factor,
the bare nucleon probability,
as a renormalization of the bare nucleon only, not as
an overall normalization  of the whole right-hand side.
In Refs. \cite{Brown,Szczurek}
the Sullivan contribution  was also multiplied with the
wave function renormalization factor.  To see the relation we note
that in quantum field theory one can write
$|N>_{\rm {phys}}=\sqrt{Z'}(|N>_{\rm{bare}}+g_{\pi NN}^0|N\pi>),$
 where $g^0$ denotes the bare coupling. The latter is related to
 the physical one (in lowest order) by $ g= \sqrt{Z'} g^0 ,$
i.e. $Z'=(1+\int dy f^{\pi /N}(y) )^{-1}. $
The prescription used in the present paper (see also Ref. 
\cite{Koepf}) is consistent with the standard nuclear physics
definition of the $\pi NN$ coupling constant derived from NN 
interaction at large distances. Of course the basic assumption 
is that we can restrict ourselves to only one pion in the air.
For the form-factors we used in present calculation (whose
conclusions are not affected by some variation of the cut-off) 
the value of $Z\approx 0.6-0.7$  suggests that this is
a good approximation. 

The  pion
light-cone momentum distribution, $f^{\pi/N}(y),$ may contain either
a nucleon or a delta final state i.e.
$f^{\pi/N}(y)= f^{\pi N/N}(y) +f^{\pi \Delta /N}(y).$ Let us first
consider
the nuclear final state. For the free nucleon the well-known result
\cite{Sullivan} for the pion distribution is
\be f^{\pi N/N}(y)= \frac{3g^2_{\pi NN}}{16\pi^2} y
\int^\infty_{M^2y^2/(1-y)} dt
\frac{|F_{\pi NN}(t)|^2t}{(t+m_\pi^2)^2}, \label{sull}
\ee
Here
$y=\frac{k_0+k_3}{M}$ gives the pion light-cone momentum
fraction, with $M$ being the physical mass of the nucleon (as an
arbitrary but convenient scale),
$F_{\pi NN} (t)$ is the $\pi NN$ vertex form-factor, while $g_{\pi NN}$ is the
$\pi^0 NN$ coupling.
The free-pion propagator, $D_\pi^0$, appears
in the above expression
in the form $(t+m^2_\pi)^{-1}$.

In the literature various prescriptions for the $\pi NN$
form-factor have been proposed.
Instead of the covariant formalism,
with the pion taken
off-shell and the final nucleon on-shell as used above, in Ref.\
\cite{MT}  and also
in work of the J\"ulich group, Ref.\ \cite{Szczurek},
old-fashioned perturbation theory in
the infinite momentum frame was used. In that formulation the pion 
and final
nucleon are taken to be on the mass shell (however, energy is not
conserved in the interaction vertex) and the form-factor is
conveniently taken
a function of the center-of-mass energy of the intermediate $\pi N$ system,
$s= (p'+k)^2 $.
In this case the probability to find a baryon in nucleon
with momentum fraction $y$ is equal
to the probabilty of finding a meson in the nucleon with momentum
fraction $1-y,$ which is not the case in the covariant approach
with the form-factor depending only on $t\equiv -k^2.$ 
 
The
impracticality of making in-medium (especially self-consistent)
calculations in the infinite-momentum frame makes us adopt the
covariant approach.
Since the covariant and the old-fashioned perturbation approach are
equivalent (apart from the effect of form factors), we assume the
conditions which should be satisfied by the first and second moments
of the two distributions for flavor-charge and momentum conservation
can be assured (at least approximately), also in the former
scheme by a suitable $p'^2$ dependence in the form-factor.
In that case (see section IV) it is enough to consider
only the change of $f^\pi(y)$ due to medium effects, which is
the main subject of the present work. Furthermore, the pion
distributions obtained in the two schemes are very similar if
in the covariant approach a monopole form-factor is used as in
our calculation.

It was confirmed \cite{Korpa93}
that dressing the pion in vacuum gives practically negligible
contribution. That is not the case, however, in nuclear medium
where delta-hole and particle-hole states play an important role.
Consequently, we want to base our calculation on the dressed-pion
self-energy and propagator, depending on $k_0$ and $k\equiv |\vec{k}|$.
The types of diagrams which contribute are shown in Fig.\ 1.
They can be summed in RPA in two steps:\\
(i) first, the pion
rescattering (without the $g'$ coupling to external nucleon)
leads to the dressed pion propagator $D_{\pi}(k_0,k)$,
$ D_{\pi}^{-1}= (D_\pi^0)^{-1}- k^2\Pi, $
where the pion self-energy $k^2\Pi$  contains both the delta-hole and
particle-hole contributions \cite{Xia1}:
\be
\Pi=\frac{\Pi_{NN}+\Pi_{N\Delta}+2 g'_{N\Delta}
\Pi_{NN}\Pi_{N\Delta}}{1-\left(g'_{N\Delta}
\right) ^2 \Pi_{NN}\Pi_{N\Delta}}. \label{pionself}
\ee
Here $\Pi_{NN}$
and $\Pi_{N\Delta}$ are the iterated particle-hole and delta-hole
self-energies divided by the common factor $k^2$
and the arguments $k_0,k$ are left out for brevity:
\begin{eqnarray}
\Pi_{NN}&=&\frac{\Pi^0_{NN}(k_0,k)}{1-g'_{NN}
\Pi^0_{NN}(k_0,k)},\label{pinn}\\
\Pi_{N\Delta}&=&\frac{\Pi^0_{N\Delta}(k_0,k)}{1-
g'_{\Delta\Delta}\Pi^0_{N\Delta}(k_0,k)},\label{pind}
\end{eqnarray}
with $\Pi^0_{NN}$ and $\Pi^0_{N\Delta}$ being the one-loop particle-hole
and delta-hole self-energies devided by $k^2$.
\\ (ii) second, the inclusion of diagrams in which the pion first
couples with $g'$ has the effect of replacing $D_\pi$ by
\be
  \tilde{D}_{\pi N}(k_0,k)= D_\pi(k_0,k)\left( 1+
\frac{X_1(k_0,k)+X_2(k_0,k)}{X_3(k_0,k)}\right)
\label{Dtilde}
\ee
with
\bea
X_1(k_0,k)&=&g'_{NN}\Pi_{NN}\left( 1+
g'_{N\Delta}\Pi_{N\Delta}\right),\\
X_2(k_0,k)&=&g'_{N\Delta}\Pi_{N\Delta}\left( 1+
g'_{N\Delta}\Pi_{NN}\right),\\
X_3(k_0,k)&=&\left[ 1-\left( g'_{N\Delta}\right)^2
\Pi_{NN}\Pi_{N\Delta}\right].
\eea
As a result in nuclear matter $D_\pi^0$ in Eq.\ (\ref{sull})
is replaced by $\tilde{D}_{\pi N}$ and in addition the integration
limits are adjusted.
The nuclear matter ground state is approximated by a mean field,
i.e. the  nucleon that emits the pion has an initial momentum
$p \le p_F$ and to satify the Pauli principle
a final momentum $p' > p_F$. The effect of
binding is included in an effective mass and a shift of energy
approximation,
i.e. $E=\sqrt{p^2+M^2_*}+c_0.$ 
The values of the parameters in the above expression
are determined from nuclear matter models and the requirement
for leading to a reasonable value of the Fermi energy.

As a result  the pion light-cone distribution (per nucleon) in isospin
symmetric nuclear matter can be expressed as
\bea
f^{\pi /A}_N(y)&=&\frac{9y  g^2_{\pi NN}M_*}{4(2\pi p_F)^3M}
\int_{-p_F}^{p_F} dp_3 \int_0^{\sqrt{p^2_F-p_3^2}}p_\perp dp_\perp
\int^{p^{\prime}_{\perp max} }_0
p'_\perp dp'_\perp \nonumber \\
 && \times\int_0^{2 \pi} d\theta k^2 F_{\pi NN}^2(k) {1\over z'}
\,|\tilde{D}_{\pi N}|^2.
\label{diagramsum}
\eea
In the above $p^{\prime}_{\perp max}= \sqrt{2z'M_* E_F-(z'^2M_*^2+M_*^2)}$,
$\theta $ is the angle between $\vec{p}_\perp$ and
$\vec{p}_\perp\!'$,
 $\vec k=\vec p-\vec{p}\,'$ (the three-momentum of the pion),
and
\begin{eqnarray}
z'&\equiv&\frac{1}{M_*}\left(-M y+p_3+\sqrt{M_*^2+p_3^2+p_\perp ^2}
\right)=\frac{1}{M_*}\left(p'_3+\sqrt{M_*^2+p'^2}\right),\\
k_0&=&\sqrt{M_*^2+p_3^2+p_\perp^2}-
\sqrt{M_*^2+p_\perp '^2+p_3'^2},
\end{eqnarray}
We note that $f^{\pi/A}(y)$ has the correct non-relativistic
low-density limit, $\mbox{lim}_{{p_F} \rightarrow 0} f^{\pi/A}(y) =
f^{\pi/N}(y),$
which is not the case in Ref.\ \cite{Bicker}.

We also consider diagrams with a delta replacing the nucleon in the final
state, in which case the relativistic analog of expression (\ref{sull})
reads
\be f^{\pi\Delta/N}(y)= \frac{g^2_{\pi N\Delta}M}{24\pi^2M^2_\Delta} y
\int dp_3 p_\perp dp_\perp {1 \over p_0}(M_+^2+t)^2
(M_-^2+t)
\frac{|F_{\pi N \Delta}(t)|^2}{(t+m_\pi^2)^2}
\rho_\Delta(p_0,p), \label{sulldelta}
\ee
where $M_\pm = M_\Delta\pm M$.
The above expression takes into account the width of the
delta through its (vacuum) spectral-function $\rho_\Delta$.
The energy of the delta is determined by the light-cone-momentum
of the pion as $p_0=M(1-y)-p_3$ and $t=p_\perp^2-M^2y^2-2Myp_3$.
We used the nonrelativistic limit of this expression, which amounts to
replacing the term $(M_+^2+t)^2(M_-^2+t)$ by the first term of low-momentum
expansion: $16 M^3 M_\Delta k^2$, where $k$ is the magnitude of the
pion's three-momentum.
In isospin-symmetric nuclear medium the pion light-cone distribution
(per nucleon) takes a form analogous to (\ref{diagramsum}):
\bea
f^{\pi /A}_\Delta(y)&=&\frac{y M g^2_{\pi N\Delta}}
{2(\pi p_F)^3M_\Delta^2}
\int_{-p_F}^{p_F} dp_3 \int_0^{\sqrt{p^2_F-p_3^2}}p_\perp dp_\perp
\int_{-\infty}^\infty dp_3'
\int^{\infty}_0
p'_\perp dp'_\perp \nonumber \\
&&\times\int_0^{2 \pi} d\theta k^2 F_{\pi N \Delta}^2(k) {1\over p'_0}
\,\rho_\Delta(p'_0,p')|\tilde{D}_{\pi \Delta}|^2,
\label{diagramsumdelta}
\eea
where
\be
\tilde{D}_{\pi \Delta}(k_0,k)=\frac{D_\pi(k_0,k)}{1-(g'_{N\Delta})
^2 \Pi_{NN} \Pi_{N\Delta}} \sum_{i=1}^3 X_{\Delta i}(k_0,k),
\ee
with
\begin{eqnarray}
X_{\Delta 1}(k_0,k)&=&g'_{N\Delta}\Pi_{NN}\left( 1+
g'_{N\Delta}\Pi_{N\Delta}\right),\\
X_{\Delta 2}(k_0,k)&=&g'_{\Delta\Delta}\Pi_{N\Delta}\left( 1+
g'_{N\Delta}\Pi_{NN}\right),\\
X_{\Delta 3}(k_0,k)&=&\left[ 1-\left( g'_{N\Delta}\right)^2
\Pi_{NN}\Pi_{N\Delta}\right].
\end{eqnarray}
In Eq.\ (\ref{diagramsumdelta})
$p'=\sqrt{p_\perp^{'2}+p_3^{'2}}$, $p'_0=
\sqrt{M_*^2+p^2}+c_0-My+p_3-p'_3$
and $k_0=\sqrt{M_*^2+p_3^2+p_\perp^2}+c_0-p'_0$.
Obviously the total pion light-cone momentum distribution
in the medium is the sum of the contributions with a nucleon
and with a $\Delta$ in the final state,
\be
f^{\pi A}(y)= f^{\pi A}_N(y) + f^{\pi A}_\Delta(y).
\ee

\section{Pion distribution and the nuclear response function}

The spin-isospin nuclear response function essentially
involves the imaginary part of the (iterated) pion self-energy
and since diagrams of the same type appear in deep inelastic
scattering off the pion emitted by the nucleon, the pion distribution
in the medium can be expressed in terms of the response function
\cite{Bicker}.

We want to clarify the relation of these two quantities for the
relativistic approach. Even though we are using nonrelativistic
nucleon and delta propagators (as well as $\pi NN$ and $\pi N\Delta$
vertices), our use of relativistic kinematics necessitates the
discussion, especially for establishing the correct low-density
limit, as well as for getting the factors of nucleon's physical
and effective mass in the expression correctly (the latter
does not appear in previous treatments \cite{Bicker,Ericson},
while in Ref.\ \cite{Brown} a peculiar rescaling is introduced).

As a first step we consider the relation of the imaginary
part of the particle-hole pion self-energy and the relevant part
of the deep inelastic diagram shown in Fig.\ 2. Iterating the
pion self-energy (with free-pion propagators in between) to get
the full response function will not change the established
relation (in the kinematic region of interest),
since the off-shellness of the emitted pion means that
cutting the free-pion propagator gives zero contribution. The square
of the absolute value of the diagram shown in Fig.\ 2b gives:
\begin{eqnarray}
L&=&-2g^2_{\pi NN} \int \frac{d^3p\,d^3p'}{(2\pi)^6 2
\sqrt{M_*^2+\vec{p}\,'^2}}
{\rm Tr}[\gamma_5(\not p_*+M_*)\gamma_5(\not p'_*+M_*)]
\Theta(p_F-|\vec p|) \Theta(|\vec p '|-p_F)\nonumber\\
&=&-4g^2_{\pi NN} \int \frac{d^3p\,d^3k}{(2\pi)^6 2 \sqrt{M_*^2+
(\vec p -\vec k)^2}}
(k^2_0-\vec{k}\,^2) \Theta(p_F-|\vec p|) \Theta(|\vec p -\vec k|-p_F),
\label{diagram}
\end{eqnarray}
where the isospin degeneracy factor of 2 is included,
$p_*$ denotes a $p$ whose zeroth component has a mean-field
shift,
$\Theta$ is the step function and $k_0\equiv \sqrt{M_*^2+\vec p \,^2}
-\sqrt{M_*^2+(\vec p - \vec k)^2}$.

On the other hand, the contribution of the mean-field nucleons to
the imaginary part of the pion self-energy in
isospin symmetric nuclear matter can be written as \cite{Phil,Korpa93}
\begin{eqnarray}
{\mbox{Im}} \Pi(k_0,k)&=&-2\times 2 g^2_{\pi NN}\pi^2
\int\frac{d^4p}{(2\pi)^4}
{\mbox{Tr}} [\gamma_5(\not p_*+M_*)\gamma_5 (\not p_*
-\not k+M_*)]
\delta((p_0-c_0)^2-\vec{p}\,^2-M_*^2) \nonumber \\
&&\times \delta((p_0-c_0-k_0)^2-(\vec{p}-\vec{k})^2
-M_*^2)\Theta(p_F-|\vec p|)\Theta(|\vec p -\vec k |-p_F) \nonumber \\
&=&-4g^2_{\pi NN}\pi \int \frac{d^3p}{(2\pi)^3 2 \sqrt{M_*^2+\vec p\,
^2}}(k_0^2-\vec{k}\,^2)
\delta((p_0-k_0)^2-(\vec{p}-\vec{k})^2
-M_*^2)\nonumber \\
&&\times \Theta(p_F-|\vec p|)\Theta(|\vec p -\vec k |-p_F).\label{impi1}
\end{eqnarray}
Integrating the above expression over $k_0$ and $\vec k$ we obtain
\bea
\int \frac{d^4k}{(2\pi)^4} \mbox{Im}\Pi(k_0,k)&=&
-2g^2_{\pi NN}\int \frac{d^3p d^3k}{(2\pi)^6 2
\sqrt{M_*^2+(\vec p -\vec k)^2}}\frac{k_0^2-\vec k\,^2}
{2\sqrt{M_*^2+\vec p \,^2}}\nonumber\\
&&\times\Theta(p_F-|\vec p|)\Theta(|\vec p -\vec k|-p_F),\label{impi}
\eea
with $k_0\equiv \sqrt{M_*^2+\vec p\,^2}-\sqrt{M_*^2+
(\vec p -\vec k)^2}$ on the right-hand side.
Comparison of expressions (\ref{diagram}) and (\ref{impi}) shows
that in general one cannot express (\ref{diagram}) in terms of
(\ref{impi}) because of presence of the momentum-dependent
factor $1/\sqrt{M_*^2+\vec p\,^2}$ in Eq.(\ref{impi}).
However,
if we approximate $\sqrt{M_*^2+\vec{p}\,^2}$
with a $p$-independent
constant (say $M_*$), which is an excellent approximation for baryon
densities not much larger than the saturation density, the
following relationship emerges:
\be
L=- 4M_* \int \frac{d^4k}{(2\pi)^4}\,{\mbox{Im}}\,\Pi(k_0,k).
\label{lapprox}
\ee
Apart from the factor $1/\sqrt{M_*^2+\vec{p}\,^2}$ whose
momentum dependence in the present context does not play an
important role,
there is
another momentum-dependent factor in expressions (\ref{diagram}) and
(\ref{impi}): $1/\sqrt{M_*^2+(\vec p -\vec k)^2}$, whose
presence is vital for correct low-density limit.
In our derivation of expression (\ref{impi1}) this factor
appears automatically since we use the fully relativistic
nucleon propagators in the particle-hole loop. However, by
using the non-relativistic limit (as, for example in Ref.\
\cite{Ericson}), this term would be replaced by $1/M_*$. This
leads then to an incorrect low-density limit of the pion distribution,
since the Sullivan expression
(\ref{sull}) for the free-nucleon is based on relativistic
calculation and contains the momentum dependence of
the mentioned term (with $\vec p =0$).

Taking into account the more complicated diagrams means
replacing $\Pi$ by $\Pi+\Pi D_\pi^0 k^2\Pi+\Pi D_\pi ^0 k^2\Pi
D_\pi^0 k^2\Pi+
\cdots = \Pi (1-D_\pi^0 k^2\Pi)^{-1}$, where $D_\pi^0$ is the free-pion
propagator. Also, when integrating over the pion momentum we
have to take into account the constraint on the pion light-cone
momentum fraction $y=(k_0+k_3)/M$. This can be imposed by an
appropriate delta-function, which should be inserted in both
expressions (\ref{diagram}) and (\ref{impi}).
Performing the $k_3$ integration and
dividing by the total number of nucleons in unit volume we
obtain the expression for the pion light-cone
momentum distribution
\be
\tilde{f}^{\pi/A} (y)=- \frac{9y M M_*}
{4\pi p^3_F} \int k_\perp dk_\perp d\omega \,\mbox{Im}
\left(\frac{k^2\Pi(\omega,k)}{1-D_\pi^0 k^2\Pi(\omega,k)}\right)
\, |D_\pi^0|^2,
\label{responsedis}
\ee
where (in the general case) the pion self-energy which contains
both the delta-hole and
particle-hole contributions is given by Eq.\ (\ref{pionself}).
In $\Pi$ and $D_\pi^0$ the argument $k$ is given by
$k=\sqrt{k_\perp^2+k_3^2}$, where $k_3=My+\omega$.
The integration region here is from zero to infinity for both
variables, since conditions on incoming nucleon (below the
Fermi sea) and outgoing nucleon (above the Fermi sea) are
automatically satisfied by the correct calculation of the imaginary
part of $\Pi$, if the nucleons are treated as a mean field.

Using the expression for the nucleon density $\rho=2p_F^3/3\pi^2$
and introducing the longitudinal response-function through
\be
R_L(\omega,k)=-\frac{1}{\pi}\frac{4MM_*}
{g_{\pi NN}^2F_{\pi NN}(k)^2}\,
\mbox{Im} \left( \frac{\Pi(\omega,k)}{1-D^0_\pi
k^2\Pi(\omega,k)}\right),
\label{responsedef}
\ee
expression (\ref{responsedis}) can be written in the usual form
\be
\tilde{f}^{\pi/A}(y)=\frac{3g_{\pi NN}^2 y}{16\pi^2 \rho}
\int dk^2 k^2 F_{\pi NN}(k)^2 \int d\omega  \frac{R_L(\omega,k)}
{(\omega^2-k^2-m_\pi^2)^2}.\label{ericsonform}
\ee
Note that the form factor and the coupling have been removed from
the imaginary part of the (iterated) pion self-energy in the
definition (\ref{responsedef}) of the response function.

We confirmed
by numerical computation that Eqs.\ (\ref{diagramsum}) and
(\ref{responsedis}) give practically identical results
(thus validating the approximation used to obtain
eq.(\ref{lapprox})),
if the pion
self-energy contains only the particle-hole contribution (assuring
the final state with only nucleon and not delta, corresponding to
diagrams summed in Eq.\ (\ref{diagramsum})).
Inclusion of the self-consistent delta-hole contribution to the
pion self-energy
takes into account also the diagrams where a delta replaces the
nucleon in the final state. This corresponds to taking into
account the contribution of diagrams summed in eq.
(\ref{diagramsumdelta}), but also some other higher-order diagrams,
which appear because the delta is itself dressed by pion-nucleon
loops (where the pion is also dressed). Numerically this shows up
as a very small difference (at values $y<0.2$) of the
distribution (\ref{ericsonform})
and the sum of eqs.(\ref{diagramsum}) and
(\ref{diagramsumdelta}). The smallnes of this difference indicates
that it is reasonable to expect that contributions of other
higher-order diagrams
left out from the calculation (for example, the pion-nucleon loop
term of the nucleon self-energy) will not affect the pion
distribution significantly.

\section{Drell-Yan scattering}

The cross-section of the Drell-Yan process $\rm{a}+\rm{b}\rightarrow
\rm{l}\bar{\rm{l}}$  can be expressed as
\be
\frac{d\sigma^{ab}}{dx_1dx_2} =
\frac{4\pi \alpha^2}{9s x_1 x_2 K(x_1,x_2)} \sum_f e^2_f \left[
 q^a_f(x_1)\bar{q}_f^b(x_2)+ \bar{q}_f^a(x_1)q^b_f(x_2) \right] ,
\label{DYcross}  \ee
where $s$ is the center-of-mass energy squared and the summation
of products of quark and antiquark distribution functions is
over flavors.
The factor $K(x_1,x_2)$
takes into account higher order QCD corrections and is of the order 1.5.
The values of $x_1,x_2$ are extracted from experiment via
the invariant mass of the lepton pair.
We are interested in the ratio of the cross-sections for
proton-nucleus and proton-deuteron scattering:
\be
R_{A/d}\equiv \frac{2}{A} \frac{d\sigma^{pA}/dx_1dx_2}{d\sigma^{pd}/dx_1dx_2},
\ee
where $A$ denotes both the nucleus and its nucleon number.

Since our in-medium computation is performed for the case
of isospin symmetric nuclear medium, we specialize to the ratio
of differential cross-sections on a nucleus consisting of equal
numbers of protons and neutrons, and on a deuteron (for which we
consider medium effects negligible)
\be
R_{A/d}=
\frac
{\sum_f e^2_f
\left\{ q^p_f(x_1)\left[ \bar{q}_f^{p/A}(x_2)+\bar{q}_f^{n/A}
(x_2)
\right] +\bar{q}_f^p(x_1)\left[ q^{p/A}_f(x_2)+q^{n/A}_f(x_2)\right]
\right\} }
{\sum_f e^2_f
\left\{ q^p_f(x_1)\left[ \bar{q}_f^p(x_2)+\bar{q}_f^n(x_2)
\right] +\bar{q}_f^p(x_1)\left[ q^p_f(x_2)+q^n_f(x_2)\right]
\right\} }.\label{ratio}
\ee
To relate the quark distribution of the bound nucleon to that
of the free nucleon we start from (\ref{pphys}), where for brevity
we write out only nucleon terms:
\bea
q_f^p(x)&=&Zq_f^{p,bare}(x)+
\frac{1}{3}\int_x^1 \frac{dy}{y} f^{N/N}(y)
\left[ q_f^{p,bare}(x/y)+2q_f^{n,bare}(x/y)\right]\nonumber\\
&&+\frac{1}{3}\int_x^1 \frac{dy}{y} f^{\pi/N}(y)
\left[ q_f^{\pi^0}(x/y)+2q_f^{\pi^+}(x/y)\right],\label{qfree}
\eea
with
\be
Z \equiv 1- \int_0^1 dy f^{N/N}(y)=1-\int_0^1 dy f^{\pi/N}(y),
\label{Z}
\ee
where the last equality in (\ref{Z}) is a requirement for flavor-charge
conservation.
Similarly, the starting expression for the quark distribution
in a nuclear proton is
\bea
\tilde{q}_f^p(x)&=&Z_A q_f^{p,bare}(x)+
\frac{1}{3}\int_x^A \frac{dy}{y} f^{N/A}(y)
\left[ q_f^{p,bare}(x/y)+2q_f^{n,bare}(x/y)\right]\nonumber\\
&&+\frac{1}{3}\int_x^1 \frac{dy}{y} f^{\pi/A}(y)
\left[ q_f^{\pi^0}(x/y)+2q_f^{\pi^+}(x/y)\right],\label{qmed}
\eea
where we specialized to the case of isospin-symmetric nuclear
matter. Adding and subtracting the expression (\ref{qfree}) to
expression (\ref{qmed}) and repeating the same procedure
for the neutron (since only the sum of these two terms is relevant
in isospin-symmetric medium) we get
\bea
\tilde{q}^p_f(x)+\tilde{q}^n_f(x)&=&
q^p_f(x)
+q^{p,bare}_f(x)\int_0^1 f^{N/N}(y)dy-\int_x^1 \frac{dy}{y}
f^{N/N}(y)q^{p,bare}_f (x/y)\nonumber\\
&&-q^{p,bare}_f(x)\int_0^A f^{N/A}(y)dy+\int_x^A \frac{dy}{y}
f^{N/A}(y)q^{p,bare}_f (x/y)+(p\rightarrow n)\nonumber\\
&&+\frac{2}{3}\int_x^A \frac{dy}{y} \left[f^{\pi/A}(y)-
f^{\pi/N}(y)\right] \left[ q_f^{\pi^0}(x/y)+
q_f^{\pi^+}(x/y)+
q_f^{\pi^-}(x/y)
\right].\label{pnmed1}
\eea
To proceed without approximation a fit of the bare structure
functions $q^{p,bare}_f(x)$ would be
required, based on expression (\ref{qfree}) and the experimentally
extracted $q^p_f(x)$. This is clearly beyond the scope of present
work and in the following we present arguments that the
expression used in Refs.\ \cite{Thomas,Llewellyn,Bick} for
the pion contribution to the change of parton distribution:
\be
\delta q^p_f(x)=\int_x^A \frac{dy}{y} \delta f^\pi (y)
q_f^\pi (x/y),\label{deltaq}
\ee
(where $\delta f^\pi(y)\equiv f^{\pi/A}(y)-f^{\pi/N}(y)$ and
isospin factors are not shown) is a good approximation for antiquark 
distributions,
the change of which is probed by the Drell-Yan pair production.

We
argue that for antiquark distributions at small $x$ it
is a good approximation to neglect the difference of the second and
third term on the right-side of Eq.\
(\ref{pnmed1}), as well as the difference of the fifth and fourth 
term. The reason is that $f^{N/N}(y)$ and
$f^{N/A}(y)$ are very small for $y<0.3$ (since the pion distributions
are negligible for $y>0.7$), thus if $x<0.3$ (the region where
antiquark distributions are significant), the zero lower limit  of
integrals can be safely shifted to $x$. Taking into account that
antiquark distributions at small $x$ behave as $1/x$
(and that $x/y$ is also small in the $y$ region of most
significant contribution), cancellation
of the considered terms follows. A much simpler argument in favor
of expression (\ref{deltaq}) is based on the assumption that
the nucleon's entire antiquark sea can be attributed to its
virtual meson cloud which, however, has not been confirmed 
\cite{Koepf}.
Since valence-quark distributions increase
with decreasing $x$ slower than $1/x$ in the small $x$ region, the
above approximation may be less good in that case, but the effect
on the Drell-Yan process of a relatively small change in quark
distribution is not significant.
The sum of the nuclear proton and neutron (anti)quark distribution
of flavor $f$ thus becomes
\be
\tilde{q}^p_f(x)+\tilde{q}^n_f(x)=
q^p_f(x)+q^n_f(x)
+\frac{2}{3}\int_x^A \frac{dy}{y} \left[ f^{\pi/A}(y)-f^{\pi/N}(y)
\right] \left[ q^{\pi^0}_f(x/y)+q^{\pi^+}_f(x/y)+q^{\pi^-}_f(x/y)
\right].
\ee
Performing a convolution on the nucleon part to take into account
the Fermi motion of the nucleons we obtain
\bea
q^{p/A}_f(x)+q^{n/A}_f(x)&=&\int_x^A \frac{dz}{z} f^N_{Fermi}(z)
\left[ q^p_f(x/z)+q^n_f(x/z) \right]
\nonumber\\
&&+
\frac{2}{3}\int_x^A \frac{dy}{y} \left[f^{\pi/A}(y)-
f^{\pi/N}(y)\right] \left[ q^f_{\pi^0}(x/y)+
q^f_{\pi^+}(x/y)+
q^f_{\pi^-}(x/y)
\right].       \label{qmedpn}
\eea

We note that the above expression satisfies flavor-charge
conservation by construction and the momentum conservation
sum-rule (with only pions included)
\be
\int_0^A z dz f^N_{Fermi}(z)+\int_0^A y dy \left[
f^{\pi/A}(y)-f^{\pi/N}(y)\right] =1.\label{momcons}
\ee
turns out to be satisfied with values of $M_*$ and $c_0$
based on nuclear matter models with accuracy better than
$1$\% (see next section).
Heavier mesons in general give much smaller contributions
than the pion \cite{MT} and would not affect significantly
the sum-rule.

One can treat the contribution of the delta in the same way, an
important difference being that even in the isospin symmetric
case the structure function of the delta appears in the structure
function of the in-medium nucleon.

To be consistent with mean-field treatment of nucleons which was
used for calculating
the pion's self-energy, for the nucleon light-cone momentum
distribution $f^N_{Fermi}$
we use the Fermi gas model with mean-field
corrections for mass and energy, $E(p)=\sqrt{M_*^2+p^2}+c_0$.
The expression, which takes into account the flux factor, is the one
obtained by Birse \cite{Birse}:
\be
f^N_{Fermi}(z)=\frac{3}{4\epsilon^3}\left[ \epsilon^2-
(z-\eta)^2 \right] \Theta \!\left( \epsilon-|z-\eta| \right),
   \label{fermirel}
\ee
where $\epsilon\equiv p_F/M, \eta\equiv (\sqrt{M_*^2+p_F^2}
+c_0)/M$
and $\Theta(x)$ is the step function.


\section{Numerical results and discussion}
First we examine the effect of Pauli blocking and Fermi motion
on the pion light-cone momentum
distribution, $f^{\pi A}(y) $ by comparing in Fig.\ 3 the free-nucleon
case (full line) to the in-medium one, but the latter calculated
with the free-pion propagator.
The pion-nucleon-nucleon
coupling has the usual value $g_{\pi NN}=13.5.$

For the actual form of the form factor various parametrizations
have been used, e.g. $F(t)= (\frac{\Lambda^2-m^2}{\Lambda^2-t})^n$
with $n=1$ (monopole) or n=2 (dipole), and also exponential forms.
The corresponding values of the cut-off, $\Lambda,$
can be related, see Ref.\ \cite{Koepf} .
Previous studies \cite{Brown,MT,Koepf,Signal,Kumano} of 
observables related to antiquark distributions obtained
cut-offs in the range $0.6-0.8$ GeV, when translated to
monopole form, which is much smaller than the one used in
the Bonn potential. We performed computations with a monopole
form using $\Lambda_{\pi NN}=0.7$ and 0.8 GeV, and also 
exponential form with 1 GeV. 
While the effect of changing the cut-off $\Lambda_{\pi NN}$
is significant for both the free nucleon case and in the medium, their
difference shows little sensitivity to this change.
This is in contrast to Ref.\ \cite{Bick} where a decrease
in the cut-off produced a decrease in the enhancement too. In the
following for all presented results we used a monopole form-factor
with $\Lambda_{\pi NN} =0.8\;$GeV, both in vacuum and in the medium.

For the values of $M_*$ and $c_0$ we turn to nuclear matter models.
While the simplest form of the Walecka model \cite{Serot} gives at
saturation density $M_*\approx 0.7$ GeV, more elaborate treatments
tend to increase the effective mass to $0.8-0.85$ GeV at saturation
density, leading to better agreement with observables
\cite{Jaminon,Horowitz}. Since we use for the Fermi momentum
$p_F=0.256\;$GeV (i.e. slightly below the saturation
density), we take $M_*=0.85\;$GeV. Adopting a smaller value
for the nucleon's effective mass decreases the in-medium
pion enhancement, as noticed in Ref.\ \cite{Brown}. The value
of the energy shift $c_0$ should be such that a reasonable
value of the Fermi energy is obtained \cite{Birse}, thus we
take $c_0=0.04\;$GeV, giving $E_F-M=-11\;$MeV. For these values
of $M_*$ and $c_0$ the momentum-conservation sum-rule 
(\ref{momcons}) is satisfied with accuracy better than $1$\% 
(for values of the $g'$ parameters in the region from 0 to 0.4), 
still leaving a little space for the effect of other mesons.

In Fig.\ 3 we see a reduction due to Pauli blocking and
practically no broadening for the effective mass of the nucleon
$M_*=0.85$ GeV (short-dashed line).
We note, however, that the decrease
of the effective mass of the nucleon contributes to the decrease
and narrowing of the pion distribution (as pointed out in Ref.\
\cite{Brown}), as can be seen also from Fig.\ 3, where the results
for $M_*=0.939$ GeV and 0.7 GeV are also shown.

For the in-medium calculation we consider first the contribution
from diagrams with a nucleon in the final state.
The results for $f^{\pi /A}_N(y)$, Eq.(\ref{diagramsum}),
for different values of $g'$ parameters are shown and
compared to the free-nucleon case in Fig.\ 4.
For $g'_{NN}$ we used the values 0, 0.4 and 0.6, while
the other two parameters are kept constant,
$g'_{N\Delta}=g'_{\Delta\Delta}=0.3$.
Increasing the $g'$ parameters in general leads to reduction
in the enhancement of pion distribution, although for small
values of $g'$ there can be some increase for certain
$y$ values, as a consequence of corresponding enhancement
in some kinematical regions \cite{Thomas}.
For the in-medium calculation we
keep the mass difference of the delta and the nucleon
the same as in free space.

We also consider the contribution from the delta in the final
state, $f^{\pi /A}_\Delta(y)$. In connection with it
we make some remarks on the  $\pi N\Delta$ coupling.
In the past the cut-off for the $\pi N\Delta $ coupling was taken
the same as for the $\pi NN$ (using SU(6)). More recently there is evidence
that the former  coupling is softer than the latter  from several sources:
\\  (i) The Juelich group has analyzed data for diffractive scattering
$p+p \rightarrow  \Delta^{++} +X $ in the one-pion exchange approximation,
 \\ (ii) Koepf et al. \cite{Koepf}  pointed out that the
 empirical  $\bar{u}-\bar{d}$ asymmetry
 can be explained in a meson cloud model only if $\Lambda_{\pi N \Delta}$
 is softer than $\Lambda_{\pi NN}$ by about 100 MeV.

In the present approach \cite{Korpa}
we have extracted $\Lambda_{\pi N \Delta}$
from a fit to $\pi N$ scattering phase-shift in the delta channel,
using three parameters:
mass of the delta, value of the coupling and cut-off, and obtained
the value for the exponential form-factor of $\Lambda_{\pi N \Delta}=0.38$
GeV (for dipole form-factor the value of the cut-off was 0.51 GeV).
This value gives an excellent fit for the phase-shift for laboratory
pion momentum up to 500 MeV. The $\pi N\Delta$ coupling used, $g_{\pi N
\Delta},$  was 20
GeV$^{-1}$ (this implies an on-shell value of 14 GeV$^{-1}$, in good
agreement with expectation based on the delta's width),
and the (bare) delta mass of 1.27 GeV.
We remark that a similarly soft pion-nucleon-delta form-factor was
extracted from pion-nucleon scattering in Ref.\ \cite{XSS}, where
also a self-consistent dressing of the pion and the delta was
performed.
Since the $\pi N \Delta$ cut-off is rather
small, we expect this contribution to be small. This is
indeed the case as one can see from Fig.\ 5. The pion distribution
is nonzero only in the region of small $y$ ($y< 0.2$), and while
for $g'=0$ (here the $g'_{\Delta\Delta}$ plays the most prominent
role) there is some enhancement over the free-nucleon case (the
full line), its value is very small. To study the effect of the
cut-off, we also performed computations with
$\Lambda_{\pi N\Delta}=0.5\;$GeV
(which gives a very poor fit to the pion-nucleon phase-shift)
and confirmed the absence of medium enhancement in that case too.
We note that using the spectral-function of the dressed
delta in the medium presents numerical difficulties, since
it is nonzero in the region where the real part of the pion
propagator takes very large values. To obtain stable results we
neglected the delta contribution if its spectral function was
smaller than 0.2 GeV$^{-1}$ (this affects the spectral-function
sum-rule negligibly).

The advantage of the expression
(\ref{responsedis}) relating the response function to the
pion light-cone-momentum distribution is that it incorporates
both contributions with a nucleon as well as delta in the final
state, if the pion self-energy contains the particle-hole and
the delta-hole contributions. Numerical evaluation confirms that
for $y>0.2$ these results agree with those based on expression
(\ref{diagramsum}), while for small $y$ there is a small
enhancement due to diagrams with the delta in the final state.
The slight difference between results based on expression
(\ref{ericsonform}) and the sum of contributions (\ref{diagramsum})
and (\ref{diagramsumdelta}) can be attributed to contributions
of higher order diagrams present in the self-consistent delta
self-energy. The results of free-space (which is approximated
by a low-density calculation with $\rho=0.1\rho_0$, where
$\rho_0$ corresponds to nucleon density of $p_F=0.256\;$GeV)
and in-medium computations,
using the same parameters as for Fig.\ 4,
are shown in Fig.\ 6.

As to the origin of the absence of significant in-medium
enhancement of the pion cloud we point to the results of 
Ref.\ \cite{Korpa} for the pion spectral function. The basic 
feature of that model is the self-consistent treatment of
the pion and the delta, with both real and imaginary parts
of the self-energies taken self-consistently into account, 
thus assuring causality for self-energies and propagators.
As a result the pion
spectral function has a main maximum whose shift to
smaller energies is much less pronounced than in previous,
less elaborate treatments, an exception being only 
Ref.\ \cite{XSS}, whose results show similar behavior
to those of Ref.\ \cite{Korpa}. 
Since this shift forms the origin
of the pion-cloud enhancement and
comes from the 
(negative) real part of the delta-hole contribution, we can 
conclude that this is where the relevant difference between
our pion's self-energy and that of previous models' rests. 
The sensitivity of the pion enhancement to this term 
requires its careful treatment, which is evidently missing in Refs.\
\cite{Thomas,Ericson}.
We emphasize once more that in Ref.\ \cite{Korpa} the real part 
of the self-energy is always calculated from its imaginary
part using a relevant dispersion relation.

In the calculation of the structure functions of nuclear nucleons
we do not take into account the delta contribution. The reason
is that we do not know the (bare) delta structure functions which
are needed as are the nucleon structure functions. The delta
structure functions are necessary even in the case of the isospin
symmetric nuclear matter. However, the above results for the pion
light-cone distribution show that there can be no noticeable
medium enhancement.

We now turn to a discussion of the DY ratio, Eq.\ (\ref{ratio}).
In applying Eq.\ (\ref{qmedpn}) we use the quark distributions
in the nucleon
and the pion from Refs.\ \cite{Eichten}, and \cite{Glueck}, respectively.
The data of Ref.\ \cite{Alde} cover a rather limited  region
of $x_2$ values around $x_2= 0.2$.
That means that in the expression (\ref{qmedpn}) the
convolution in the first term on the right-hand side
is relatively unimportant and that in the integral in the second term rather
large $y$ values are sampled, typically $y \geq 0.3$ and larger.
Therefore one probes  rather large 3-momenta
in the  response function $D(k_0,k)$, Eq.\ (\ref{Dtilde}).
From $k^2> k_3^2 $ and $k_3= -k_0+ my $ one has  $ |k| > |k_0| + my$ ,
i.e.\ typically $k> 400 $MeV/$c$.
As a consequence Pauli blocking (which is relevant for small $k$ )
is not very effective;
also the delta-hole component at $k_{0,\Delta}$
is relatively unimportant since it involves still higher momenta,
$k > my+|k_{0,\Delta}|$.
This is rather different from the situation for
the longitudinal spin response function
probed in e.g. the quasi-free (p,n) reaction.

Results for the ratio (\ref{ratio}) of Drell-Yan cross sections are
shown in Fig.\ 7, for different values of the Migdal $g'$ parameters.
A general feature of all these plots is the decreasing trend for
larger values of $x_2$ ($x_2>0.3$), leading to values below
unity. The enhancement for smaller values of $x_2$ even with
no Migdal correction is quite modest.
Values of the $g'_{NN}$ around 0.3 lead to practically no enhancement.

Compared to Ref.\ \cite{Brown} we still obtain a net medium
DY enhancement by $10\%$
for $g'<0.4$ in the region $0.1 <x_2  <0.3$.
This difference can be ascribed to the the fact that
in \cite{Brown} an {\em ad hoc} $Z$-factor (see discussion
below Eq.\ (\ref{pphys}))
occurs on right-hand side of Eq.\ (\ref{qmedpn}) which
is not present in our approach. In addition
in \cite{Brown} the assumed density dependent in-medium reduction
of the cut-off $\Lambda$ in the $\pi NN$  form factor  appears the
main mechanism
to reduce the pionic enhancement. However we feel there is no
justification for such a medium vertex renormalization.

For direct comparison with measured results of Ref.\ \cite{Alde} we
calculated the ratio of nuclear and nucleon cross-section for
given values of $x_2$ and the condition $x_1 > x_2 +0.2$,
corresponding
to the experimental cut-off (here we assume that the factor
$K(x_1, x_2)$
is constant). In accordance with the above discussion we verified
numerically that the ratio in the given $x_2$ region is mainly
sensitive to the pion distribution in the region $y>0.3$. Only the
region of $x_2$ around 0.3 shows some sensitivity to smaller $y$
values. This means that the drop of the ratio below one for
$x_2$ around 0.05 (if the effect is real), probably cannot be
explained by a change of the pion distribution.

Based on our results we can conclude that the treatment of the delta,
although seemingly of secondary importance because of the kinematical
region of the delta-hole contribution, does play a significant
role. Including both the real and imaginary parts of its self-energy
in a self-consistent way eliminates a large part of the
pion-distribution in-medium enhancement and leads to agreement with
experimental results on Drell-Yan scattering for modest values
of the $g'$ parameters. The reduction in the in-medium pion
enhancement is achieved without intruducing a medium reduction of
the pion-nucleon-nucleon vertex cut-off parameter or a
renormalization factor to suppress the pion-cloud contribution.

Finally we note that in the present paper we have restricted
ourselves to isospin symmetric nuclear matter. On the other
hand some of the data involve nuclei like Fe and W,
which have a neutron excess.
It is an interesting question whether there are additional
medium effects to the asymmetry  $\bar{u}(x)-\bar{d}(x)$ in case of
neutron excess. We plan to address this question in the
future.

\acknowledgments
This research was supported in part by an NWO (Netherlands)
fellowship and the
Hungarian Research Foundation (OTKA) grant T16594.


\begin{figure}
\caption{Types of diagrams involving the pion self-energy
and dressed propagator (dashed line), used for calculating the pion
distribution in the medium. The double line denotes either a nucleon
or delta.}
\label{fig1}
\end{figure}

\begin{figure}
\caption{The particle-hole pion self-energy (a), and the
pion emission part of the diagram showing the deep-inelastic
scattering off the nucleon's pion cloud (b).}
\label{fig2}
\end{figure}

\begin{figure}
\caption{Effect of Pauli blocking and Fermi motion on the pion
distribution. The full line corresponds to to the free nucleon,
the short-dashed line to nuclear matter with $p_F=0.256$ GeV
and $M_*=0.85$ GeV, but
with free-pion propagator. The long-dashed line corresponds to
$M_*=0.7$ GeV, while the dot-dashed line is for $M_*=0.939$ GeV.}
\label{fig3}
\end{figure}

\begin{figure}
\caption{Pion distribution in the nuclear medium with Fermi momentum
$p_F=0.256$ GeV and for the free nucleon.
Full line is for the free nucleon, long dashed line for medium with
$g'_{NN}=0,\;g'_{N\Delta}=g'_{\Delta\Delta}=0.3$,
short dashed for
$g'_{NN}=0.4,\;g'_{N\Delta}=g'_{\Delta\Delta}=0.3$
and dot-dashed for
$g'_{NN}=0.6,\;g'_{N\Delta}=g'_{\Delta\Delta}=0.3$.
}
\label{fig4}
\end{figure}

\begin{figure}
\caption{Delta contribution to the pion distribution in the nuclear
medium with Fermi momentum
$p_F=0.256$ GeV and for the free nucleon.
Full line is for the free nucleon, long dashed line for medium with
$g'_{NN}=g'_{N\Delta}=g'_{\Delta\Delta}=0$,
short dashed for
$g'_{NN}=g'_{N\Delta}=g'_{\Delta\Delta}=0.3$
and dot-dashed for
$g'_{NN}=g'_{N\Delta}=g'_{\Delta\Delta}=0.5$.
The effective mass of the delta in the medium is $M^*_\Delta=
1.18$ GeV (assuring the same effective mass difference for the
delta and nucleon in the medium as in the free space),
and the $\pi N\Delta$ vertex cut-off $\Lambda_{\pi N\Delta}=0.38$ GeV.
}
\label{fig5}
\end{figure}

\begin{figure}
\caption{Pion distribution in the nuclear medium, based on computation
using the response-function approach, expression
(\protect{\ref{ericsonform}}). Full line is for the
low-density case with $p_F=0.119\;$GeV, corresponding to 1/10 of the
density used for in-medium calculations.
Long dashed line is for medium with
$g'_{NN}=g'_{N\Delta}=g'_{\Delta\Delta}=0$,
short dashed for
$g'_{NN}=0.4, g'_{N\Delta}=g'_{\Delta\Delta}=0.3$
and dot-dashed for
$g'_{NN}=0.6, g'_{N\Delta}=g'_{\Delta\Delta}=0.3$.
}
\label{fig6}
\end{figure}

\begin{figure}
\caption{Ratio (\protect{\ref{ratio}}) of the Drell-Yan
cross-sections.
Full line is for $x_1=0.3$, dashed line for $x_1=0.4$ and
dot-dashed line for $x_2=0.5$.
The Migdal parameters are:
a) $g'_{NN}=g'_{N\Delta}=g'_{\Delta\Delta}=0$,
b) $g'_{NN}=0.4, g'_{N\Delta}=g'_{\Delta\Delta}=0.3$.
}
\label{fig7}
\end{figure}

\begin{figure}
\caption{Experimental results from Ref.\ \protect{\cite{Alde}} compared
to our calculation for different $g'$ values. Full line is for
$g'_{NN}=g'_{N\Delta}=g'_{\Delta\Delta}=0.4$, dashed line for
$g'_{NN}=0.4,\,g'_{N\Delta}=g'_{\Delta\Delta}=0.3$, dot-dashed line for
$g'_{NN}=0.6,\,g'_{N\Delta}=g'_{\Delta\Delta}=0.3$.
}
\label{fig8}
\end{figure}


\newpage
\vspace{10mm}
\epsfxsize=15cm
\centerline{\epsffile{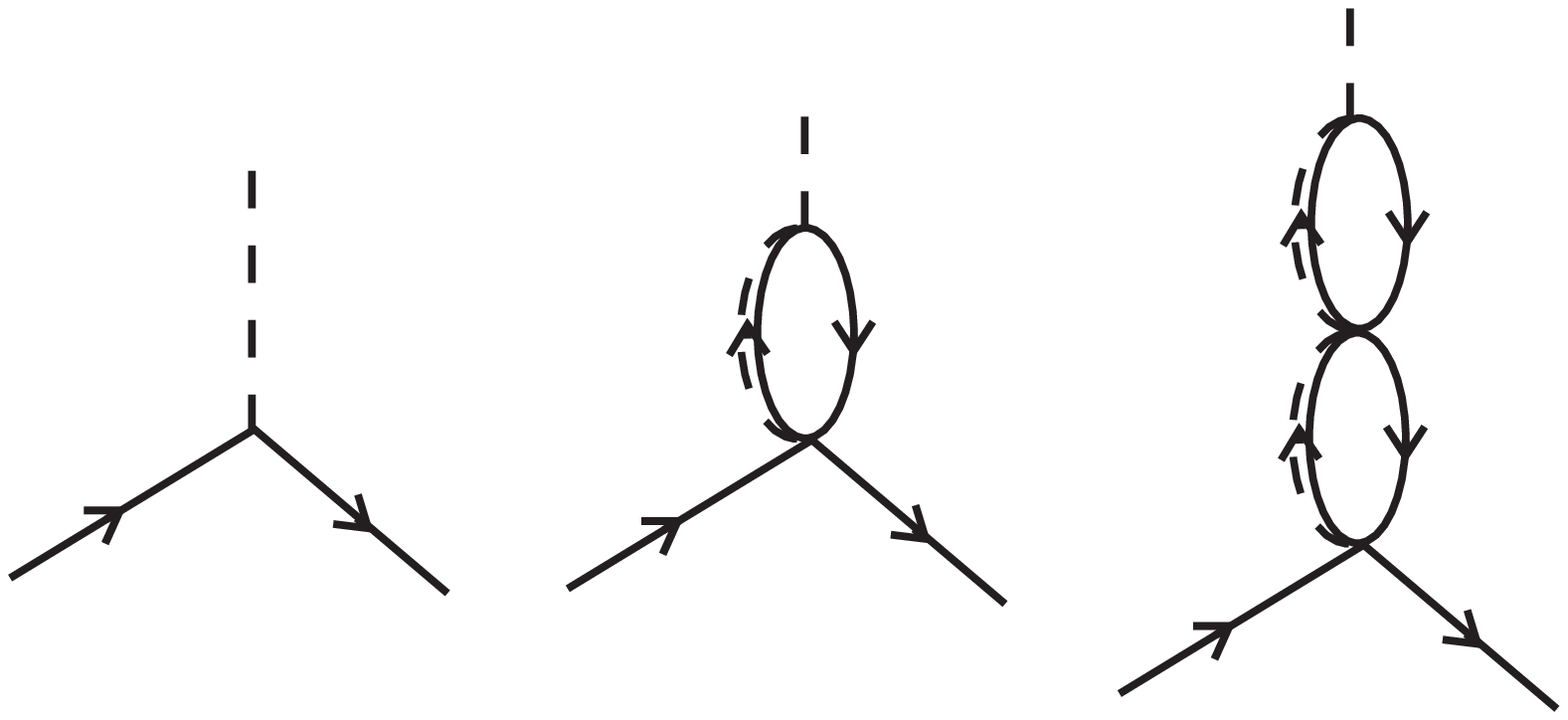}}
\begin{center}
Fig.\ 1
\end{center}

\newpage
\vspace{10mm}
\epsfxsize=14cm
\centerline{\epsffile{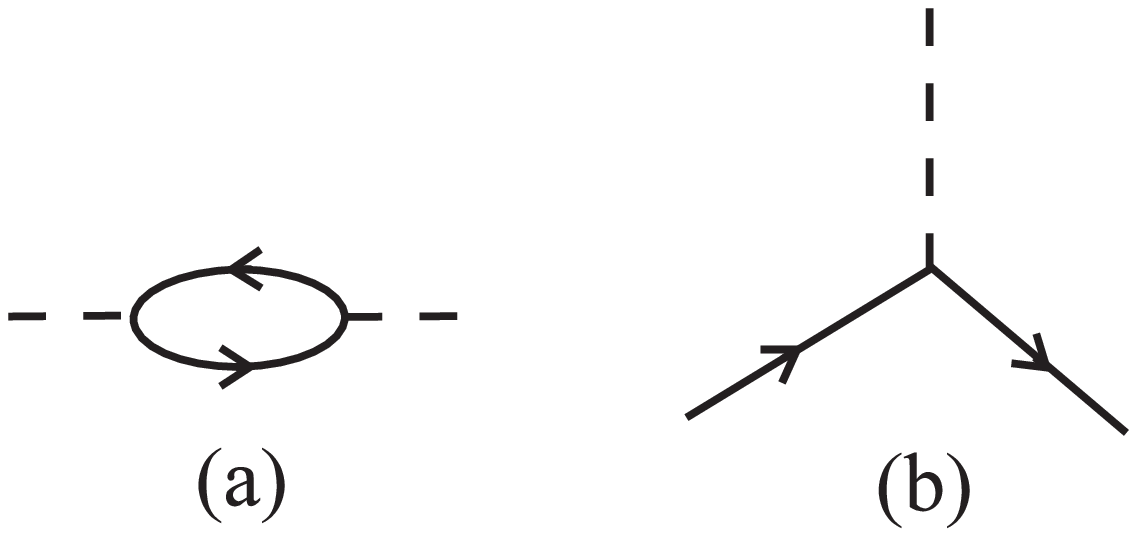}}
\begin{center}
Fig.\ 2
\end{center}

\newpage
\vspace{10mm}
\epsfxsize=14cm
\centerline{\epsffile{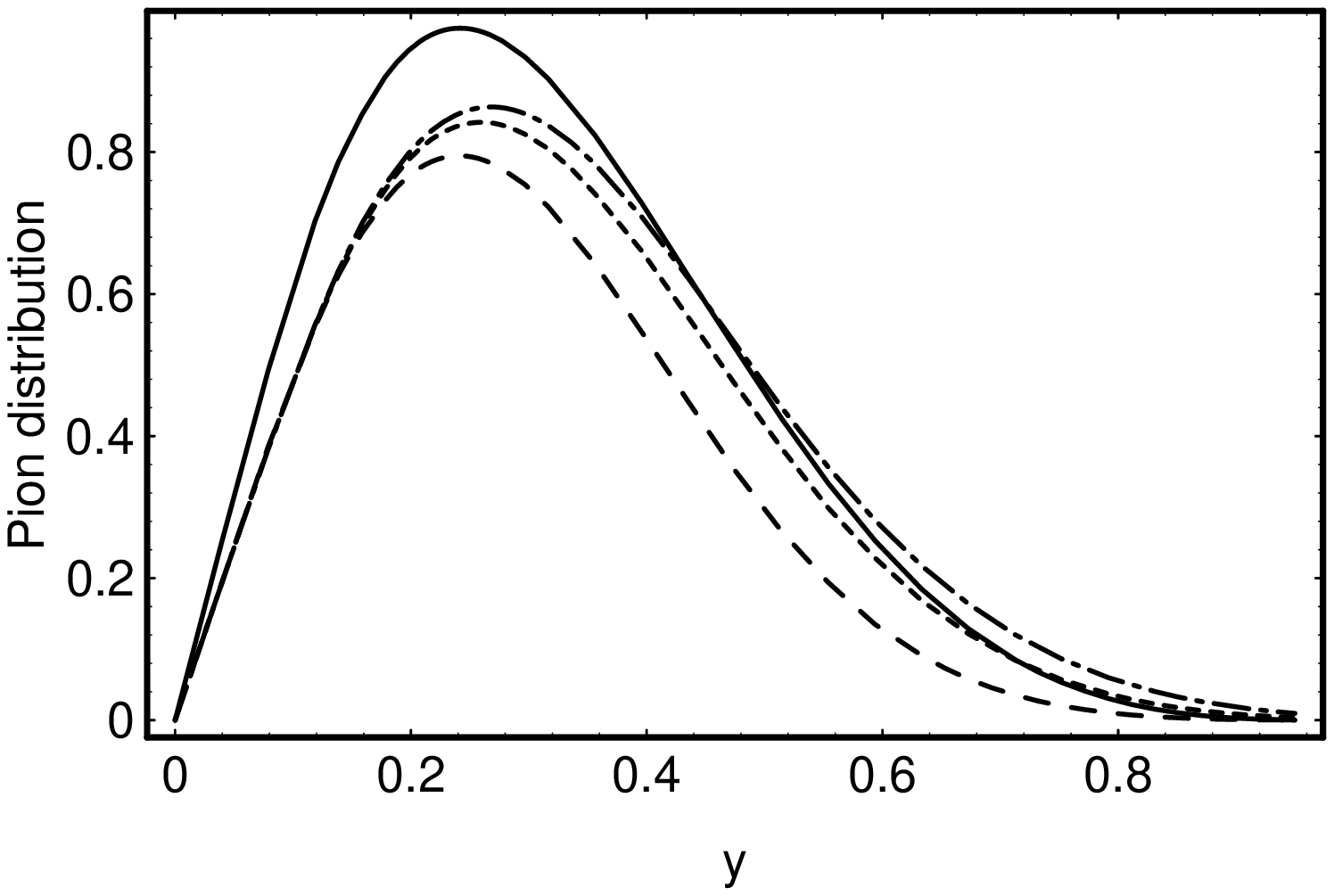}}
\begin{center}
Fig.\ 3
\end{center}

\newpage
\vspace{10mm}
\epsfxsize=14cm
\centerline{\epsffile{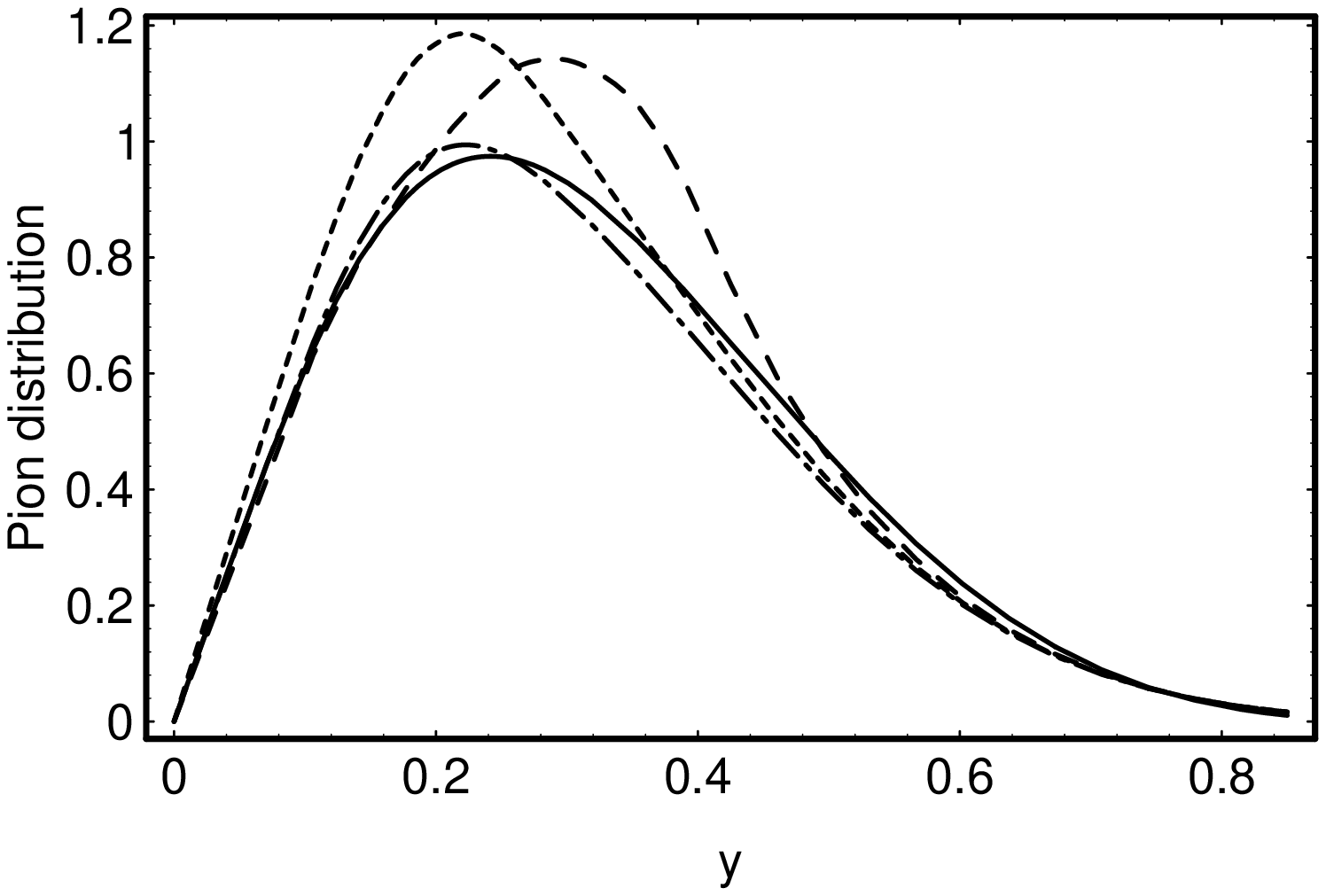}}
\begin{center}
Fig.\ 4
\end{center}

\newpage
\vspace{10mm}
\epsfxsize=14cm
\centerline{\epsffile{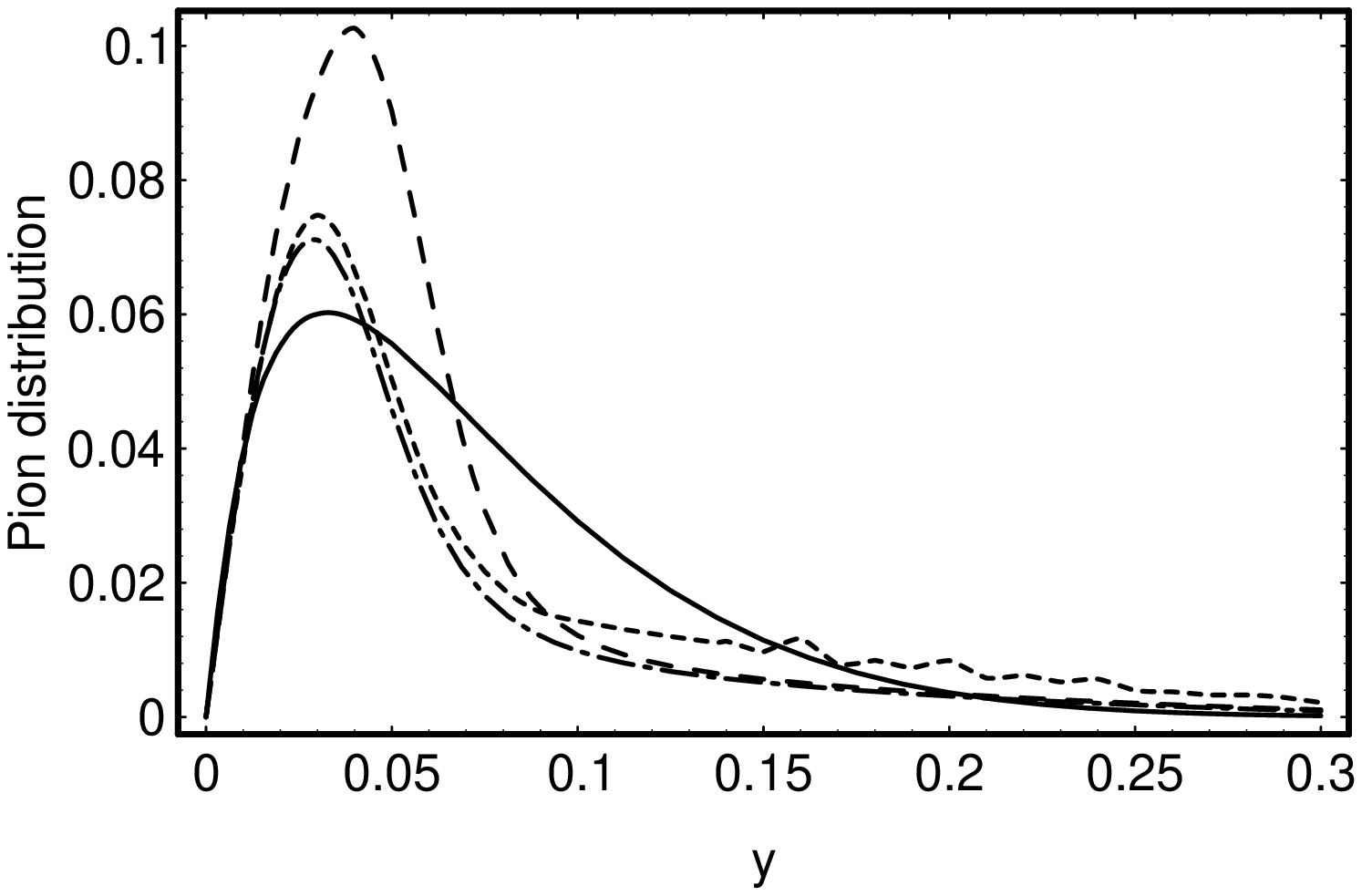}}
\begin{center}
Fig.\ 5
\end{center}

\newpage
\vspace{10mm}
\epsfxsize=14cm
\centerline{\epsffile{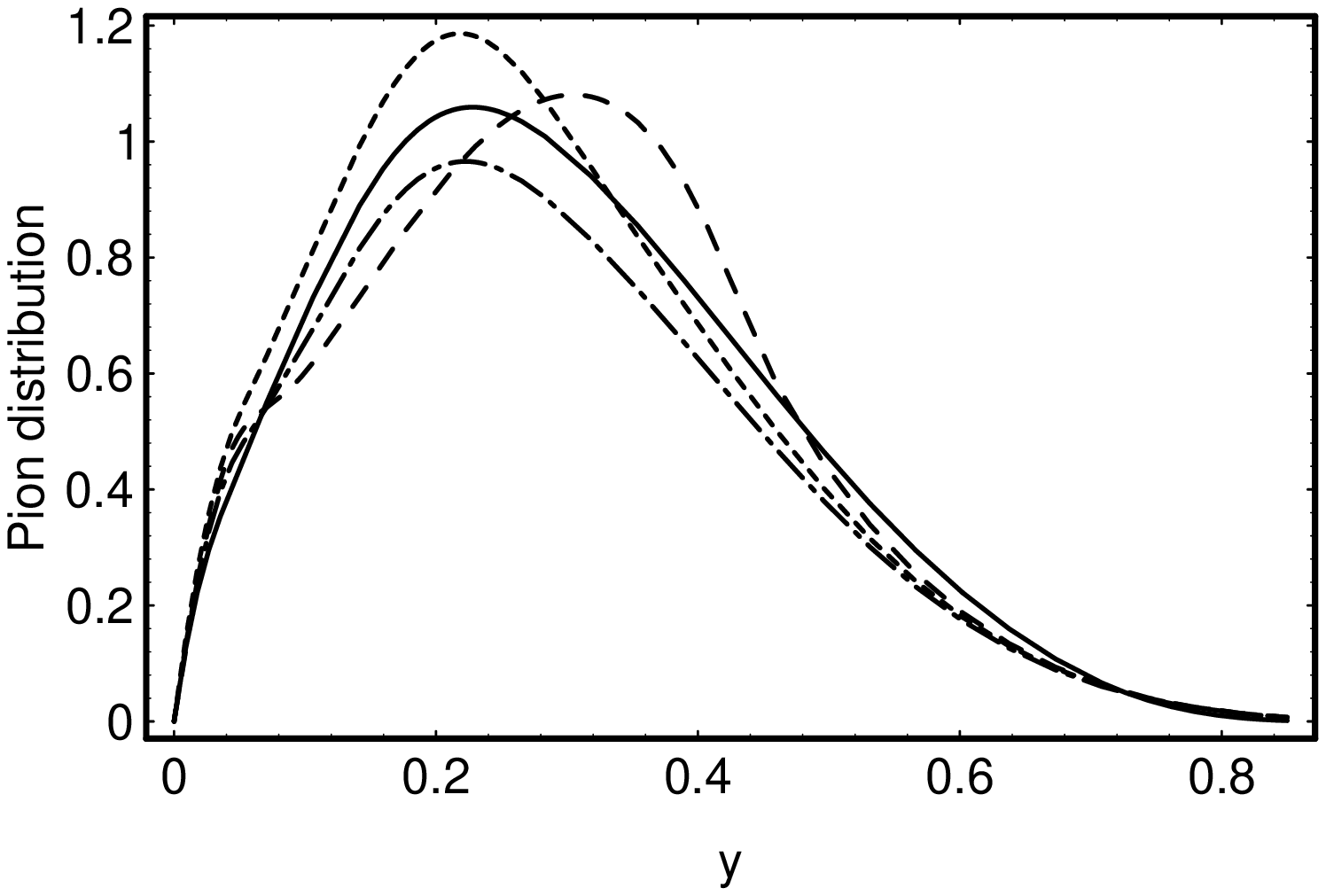}}
\begin{center}
Fig.\ 6
\end{center}

\newpage
\vspace{10mm}
\epsfxsize=14cm
\centerline{\epsffile{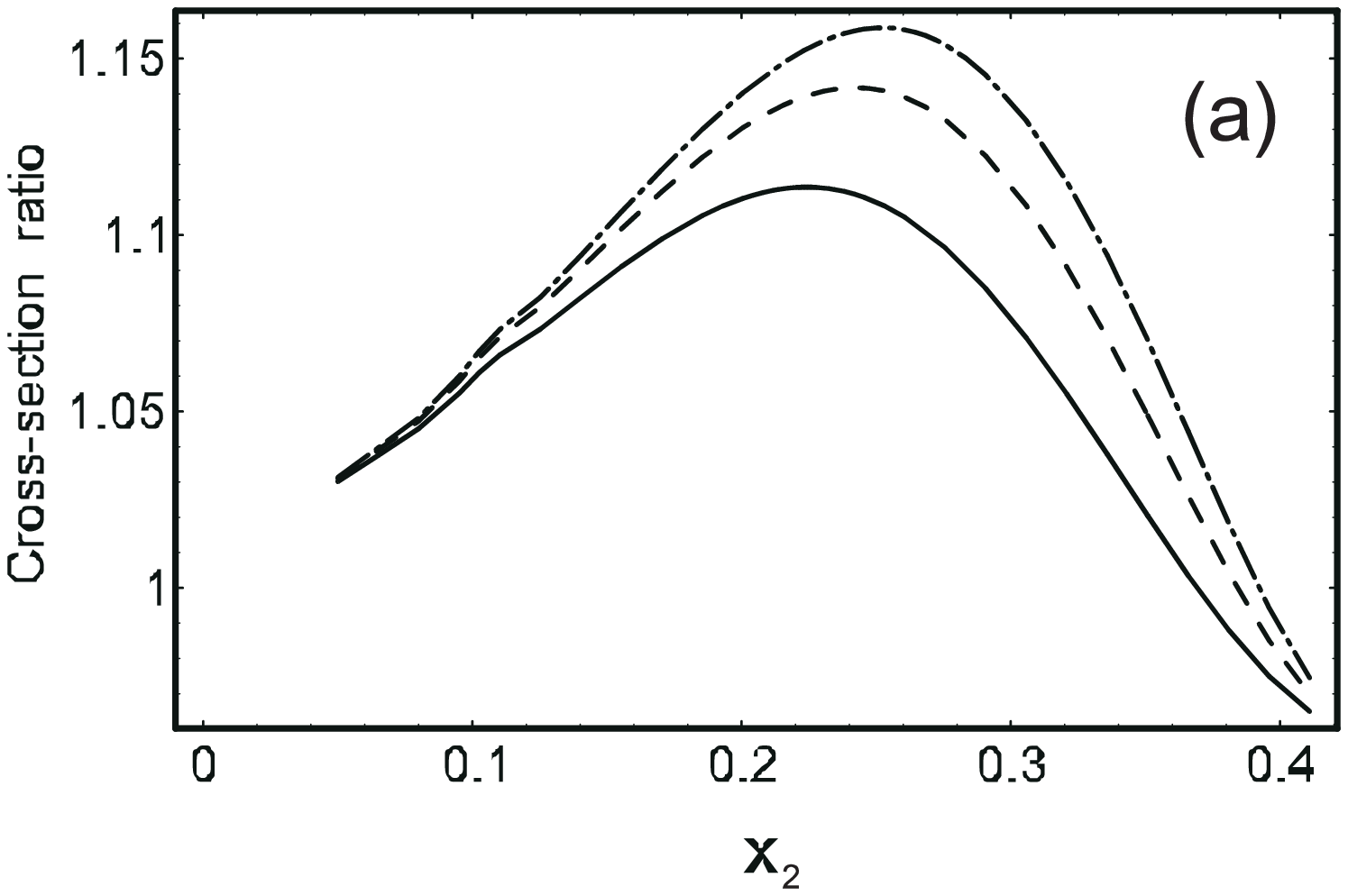}}
\begin{center}
Fig.\ 7a
\end{center}

\newpage
\vspace{10mm}
\epsfxsize=14cm
\centerline{\epsffile{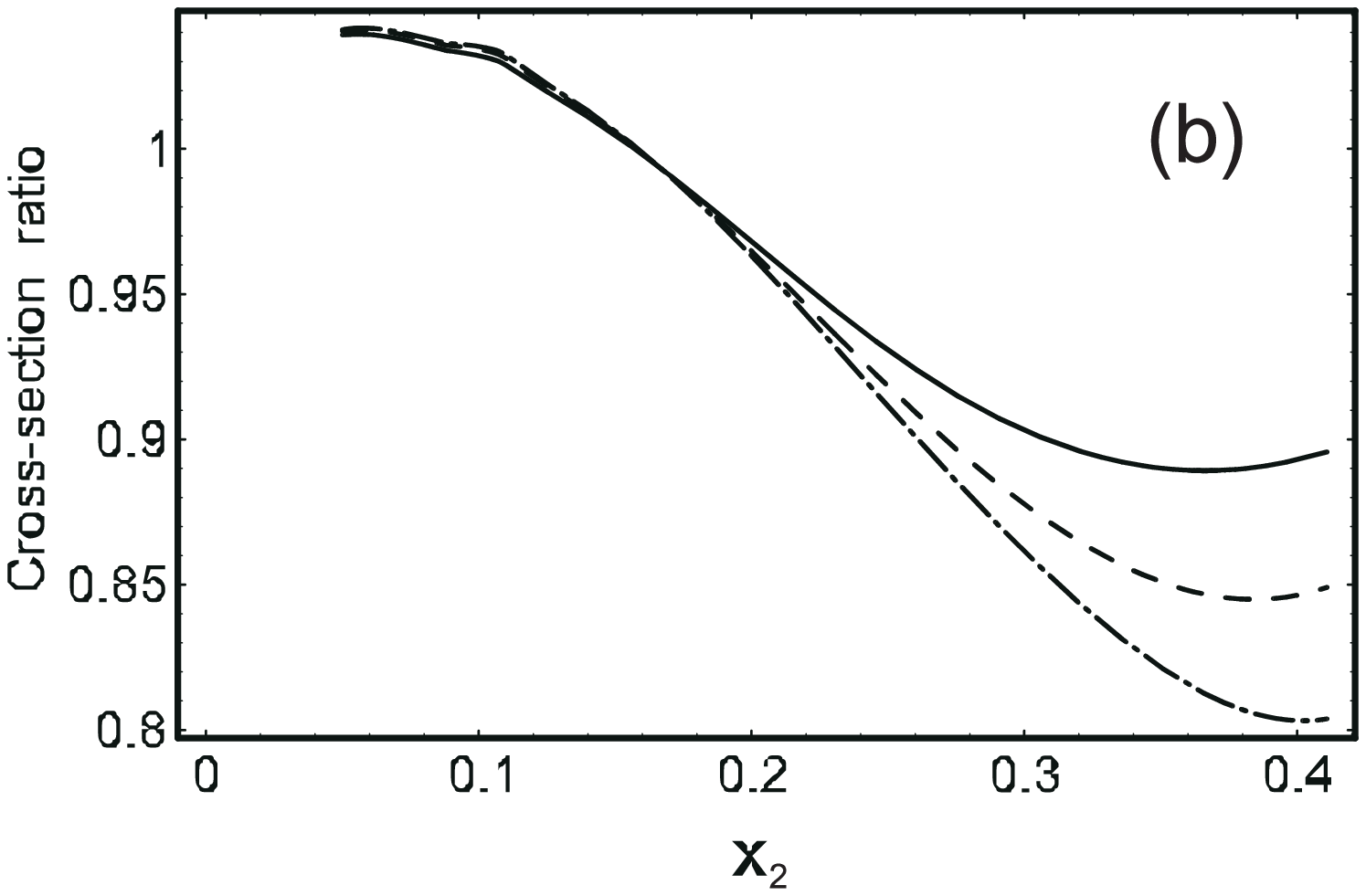}}
\begin{center}
Fig.\ 7b
\end{center}

\newpage
\vspace{10mm}
\epsfxsize=14cm
\centerline{\epsffile{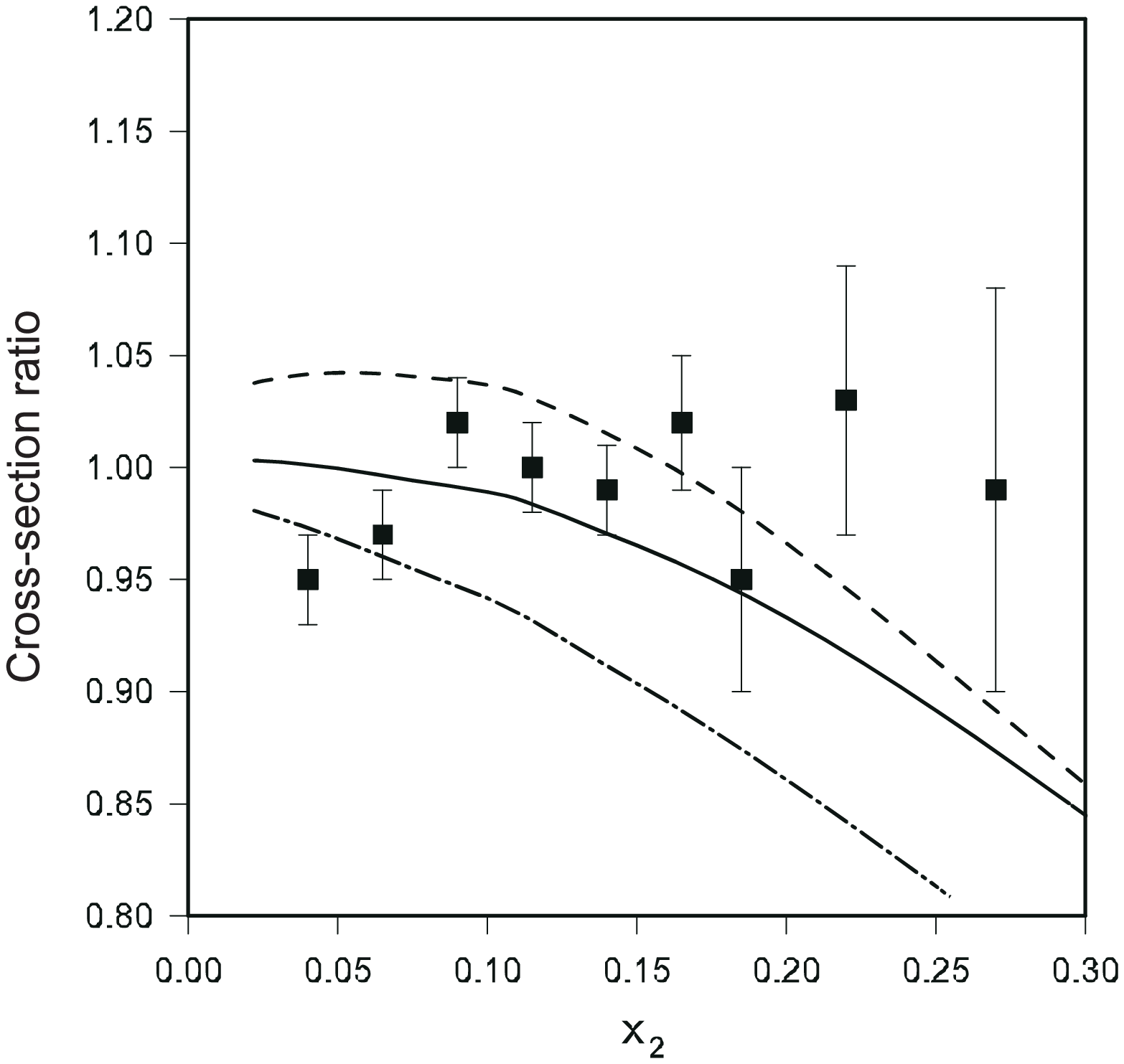}}
\begin{center}
Fig.\ 8
\end{center}

\end{document}